\documentclass[superscriptaddress,showkeys,useAMS,twocolumn,nofootinbib]{revtex4-1}
\pdfoutput=1
\usepackage{graphics, graphicx, amsmath, amssymb}
\usepackage{epstopdf}
\usepackage{dsfont}
\usepackage{hyperref}
\usepackage{color}

\newcommand{\eq}[1]{\begin{equation}\begin{aligned}#1\end{aligned}\end{equation}}
\newcommand{\iu}{\text{i}}
\newcommand{\eu}{\text{e}}
\newcommand{\ha}{\hat{a}}
\newcommand{\had}{\hat{a}^\dagger}
\newcommand{\hb}{\hat{b}}
\newcommand{\hbd}{\hat{b}^\dagger}
\newcommand{\ket}[1]{\left|#1\right\rangle}
\newcommand{\bra}[1]{\left\langle#1\right|}

\begin{document}

\title{Perfect polarization for arbitrary light beams}
\date{\today}
\author{Aaron Z. Goldberg}
\email{goldberg@physics.utoronto.ca}
\affiliation{Department of Physics, University of Toronto, Toronto, ON, M5S 1A7}
\author{Daniel F. V. James}
\affiliation{Department of Physics, University of Toronto, Toronto, ON, M5S 1A7}
\begin{abstract}
	Polarization of light is harnessed in an abundance of classical and quantum applications. Characterizing polarization in a classical sense is done resoundingly successfully using the Stokes parameters, and numerous proposals offer new quantum counterparts of this characterization.
	The latter often rely on distance measures from completely polarized or unpolarized light. We here show that the accepted class of perfectly polarized quantum states of light is severely lacking in terms of both pure states and mixed states. By appealing to symmetry and geometry arguments we determine all of the states corresponding to perfect polarization, and show that the accepted class of completely polarized quantum states is only a subset of our result. We use this result to reinterpret the canonical degree of polarization, commenting on its interpretation for classical and quantum light. Our results are necessary for any further characterizations of light's polarization.
\end{abstract}
\maketitle

\section{Introduction}
Polarization is a fundamental property of light \cite{MandelWolf1995, BornWolf1999} (see \cite{Luis2016} for a recent review). It is readily manipulated \cite{ChoquetteLeibenguth1994,Grier2003,Straufetal2007} and measured \cite{Hauge1976,Mishchenkoetal2011}, making it integral to fields such as communications \cite{HanLi2005}, remote sensing \cite{Deuzeetal2001,Schott2009}, weather radar \cite{RicoRamirezCluckie2008}, and astrophysics \cite{Dulketal1994}. 
Polarization is well-defined quantum mechanically, and is the parameter of choice in many quantum optical and quantum information protocols including quantum key distribution \cite{Bennettetal1992, Mulleretal1993}, EPR tests \cite{Kwiatetal1995}, and quantum tomography \cite{Jamesetal2001}. However, quantum mechanical characterizations of polarization are incomplete.

In classical physics, polarization is completely characterized by the Stokes parameters \cite{Jackson1999,BornWolf1999}. 
Deterministic beams of light are said to be ``perfectly" polarized, as are ensemble-averaged beams whose Stokes parameters obey $S_1^2+S_2^2+S_3^2=S_0^2$ \cite{BornWolf1999}. Natural light,\footnote{I.e., stochastic light. This includes incandescent sources such as light bulbs and sunlight. Solar radiation that reaches the Earth's surface, however, is partially polarized due to scattering with atmospheric particles \cite{Hariharan1969}.} on the other hand, is completely unpolarized, with Stokes parameters obeying $S_1=S_2=S_3=0$ \cite{Wolf2007}. All statistically stationary fields can be uniquely decomposed into the sum of a perfectly polarized and a perfectly unpolarized beam of light, and the degree of polarization quantifies the relative contribution of these two beams \cite{BornWolf1999}. We here investigate the completeness of this description for quantum states of light.

It has been established that quantum fluctuations give rise to phenomena not described by the Stokes parameters \cite{PrakashChandra1971,Klyshko1992,Bushevetal2001,Luis2002,AgarwalChaturvedi2003,Bjorketal2015,LuisDonoso2016,ShabbirBjork2016}, 
leading to a strong push in recent years to find new methods of characterizing the degree of polarization from a quantum mechanical perspective \cite{AlodjantsArakelian1999,Luis2002,Luis2007OptComm,Klimovetal2010, Bjorketal2010}. 
Perfectly unpolarized quantum states of light are well-understood both with the Stokes parameters \cite{JauchRohrlich1955,MandelWolf1995,BornWolf1999} and without \cite{PrakashChandra1971,Agarwal1971,Lehneretal1996,Soderholmetal2001}.
The same cannot be said of perfectly polarized quantum states; this is the focus of the present work. 

In this paper we characterize all of the perfectly polarized states of light. This was originally done by Mehta and Sharma in 1974 \cite{MehtaSharma1974}, but we show that their result was incomplete. We generalize the known results for both pure and mixed states, and phrase the former in terms of the geometrically intuitive SU(2) coherent states and the Majorana representation. Our list of states with perfect polarization is exhaustive. We then use this result to explain the canonical definition of degree of polarization, thereby determining exactly how much information is contained in the readily-measured Stokes parameters.

Our investigation of perfect polarization has many experimental consequences. Perfectly polarized light is fully transmitted through an ideal polarizer \cite{Collett1992}. 
It can be used for polarization holography \cite{NikolovaRamanujam2009}, communications, ellipsometry, the electro-optical effect, and more \cite{Collett1992,Huard1997}.
Polarization is related to indistinguishability and entanglement, with perfectly polarized light embodying the largest amount of separability and thus the most knowledge of the state \cite{Mandel1991,QianEberly2011,deZela2014}; highly polarized beams are necessary for precise experiments. It is fundamental to understand such ubiquitously used properties of light.

This work is organized as follows. In Section \ref{sec:classical quantum polarization} we review polarization from classical and quantum standpoints, including a discussion of non-canonical definitions of degree of polarization. Sections \ref{sec:fixed particle number} and \ref{sec:non fixed particle number} investigate perfectly polarized quantum states for fixed and indeterminate particle number, respectively. The former gives a geometrical interpretation to the currently accepted class of perfectly polarized pure states,  and the latter shows that our new results generalize all of the previously-accepted classes of perfectly polarized quantum states. In Section \ref{sec:decomposition into polarized and unpolarized} we comment on what our results imply for the canonical definition of degree of polarization, showing that the decomposition of a state into perfectly polarized and unpolarized components is no longer unique; and, in Section \ref{sec:conclusions}, we conclude.

\section{Classical and Quantum Polarization}
\label{sec:classical quantum polarization}
\subsection{Stokes parameters and degree of polarization}
The most general description of a plane wave propagating in direction $\vec{k}$ is given by the electric field
\eq{
	\vec{E}\left(\vec{x},t\right)=\left(\hat{\epsilon}_a\alpha+\hat{\epsilon}_b\beta\right)\eu^{\iu \vec{k}\cdot\vec{x}-\iu\omega t},
	\label{eq:plane wave E field}
}
for constants $\alpha$ and $\beta$ and orthogonal polarization vectors $\hat{\epsilon}_a$ and $\hat{\epsilon}_b$. This can be characterized by the four Stokes parameters
\eq{
	S_{0,\text{pw}}&=\left|\hat{\epsilon}_a\cdot\vec{E}\right|^2+\left|\hat{\epsilon}_b\cdot\vec{E}\right|^2=\left|\alpha\right|^2+\left|\beta\right|^2\\
	S_{1,\text{pw}}&=\left|\hat{\epsilon}_a\cdot\vec{E}\right|^2-\left|\hat{\epsilon}_b\cdot\vec{E}\right|^2=\left(\left|\alpha\right|^2-\left|\beta\right|^2\right)\\
	S_{2,\text{pw}}&=2\Re\left[\left(\hat{\epsilon}_a\cdot\vec{E}\right)^*\left(\hat{\epsilon}_b\cdot\vec{E}\right)\right]=\left(\alpha^*\beta+\alpha\beta^*\right)\\
	S_{3,\text{pw}}&=2\Im\left[\left(\hat{\epsilon}_a\cdot\vec{E}\right)^*\left(\hat{\epsilon}_b\cdot\vec{E}\right)\right]=-\iu\left(\alpha^*\beta-\alpha\beta^*\right).
	\label{eq:classical Stokes parameters}
}
The Stokes parameters for a plane wave associated with Eq. (\ref{eq:plane wave E field}) satisfy the identity $S_{0,\text{pw}}^2=S_{1,\text{pw}}^2+S_{2,\text{pw}}^2+S_{3,\text{pw}}^2$ 
such that the vector $\vec{S}_{\text{pw}}\equiv\left(S_{1,\text{pw}},S_{2,\text{pw}},S_{3,\text{pw}}\right)$ normalized by $S_0$ spans a unit sphere known as the Poincar\'e sphere \cite{Collett1992}. 

Stochastic light requires taking time or ensemble averages of Eq. (\ref{eq:classical Stokes parameters}) $S_\mu=\left\langle S_{\mu,\text{pw}}\right\rangle_\text{classical}$, with the vector $\vec{S}/S_0$ in general lying inside of the Poincar\'e sphere, and so we define the degree of polarization as per Wiener
\footnote{Refs. \cite{BillingsLand1948,Walker1954, Wolf1954} also allude to this definition of degree of polarization; however, the first to use the Stokes parameters explicitly in the definition after they were rediscovered by Soleillet in 1929 \cite{Soleillet1929} and made famous by Chandrasekhar in the 1940s \cite{Chandrasekhar1950} was Wolf in 1959 \cite{Wolf1959} (see Refs. \cite{JauchRohrlich1955,Collett1992} for detailed historical discussions).} 
by the formula \cite{Wiener1930} 
\eq{p=\frac{\left|\vec{S}\right|}{S_0}\label{eq:p degree of polarization}.} 
Only perfectly polarized states of light have $p=1$; in general, we have the relation $p\leq 1$ \cite{Jackson1999,BornWolf1999}. Any classical beam of light can be written as the unique sum of a perfectly polarized and a completely unpolarized beam, and $p$ quantifies the relative contributions of the two \cite{Wolf1959,BornWolf1999,Collett1992}.

When we consider quantum fluctuations, $\vec{E}$ becomes an operator with $\alpha\to\sqrt{\frac{\hbar\omega}{2V\epsilon_0}}\hat{a}$ and $\beta\to\sqrt{\frac{\hbar\omega}{2V\epsilon_0}}\hat{b}$ inside a quantization volume $V$, where the latter obey bosonic commutation relations $\left[\ha_i,\had_j\right]=\delta_{ij},\,\ha_i\in\left\{\ha,\hb\right\}$. $\ha$ and $\hb$ can represent annihilation operators for any two orthogonal polarizations of light. The Stokes parameters are normalized by $\left|\alpha\right|^2+\left|\beta\right|^2$ and promoted to operators defined by \cite{Fano1949,Fano1954,JauchRohrlich1955} \eq{
	&\hat{S}_0=\had\ha+\hbd\hb \\
	&\hat{S}_1=\had\ha-\hbd\hb \\
	&\hat{S}_2=\had\hb+\hbd\ha \\
	&\hat{S}_3=-\iu\left(\had\hb-\hbd\ha\right)
	\label{eq:Quantum Stokes operators}.
	}
These operators satisfy the $\mathfrak{su}$(2) algebra
\eq{
	\left[\hat{S}_\mu,\hat{S}_\nu\right]&=2\iu\left(1-\delta_{\mu 0}\right)\left(1-\delta_{\nu 0}\right)\sum_{j=1}^3\epsilon_{\mu\nu j} \hat{S}_j\\
	\hat{S}_1^2&+\hat{S}_2^2+\hat{S}_3^3=\hat{S}_0^2+2\hat{S}_0,
}
and are related to the classical Stokes parameters by $S_\mu=\left\langle\hat{S}_\mu\right\rangle$, where $\left\langle\bullet\right\rangle$ denotes the quantum expectation value
\cite{JauchRohrlich1955,Collett1970}.

Recent work has made significant progress in characterizing higher-order moments than the Stokes parameters for situations in which $S_1=S_2=S_3=0$
but ``hidden" polarization still exists. 
These are states that can be decomposed into two arbitrary, orthogonal, plane-polarized components whose phases are not statistically independent; i.e., $\left\langle\had\vphantom{a}^n\ha^m\right\rangle \neq 0$ for some $n,m \in \mathds{N}$ \cite{PrakashChandra1971,Klyshko1992}. Here we focus on conventional polarization, with an emphasis on the true quantum regime that allows for indeterminate numbers of photons.

\subsection{Alternative definitions for degree of polarization}
Many authors have expressed misgivings regarding the definition of degree of polarization in Eq. (\ref{eq:p degree of polarization}) \cite{Klyshko1992,AlodjantsArakelian1999,Luis2002,Luis2007OptComm,Klimovetal2005,Klimovetal2010, Bjorketal2010}. One of the first challenges to Eq. (\ref{eq:p degree of polarization}) is the inability to simultaneously measure all three components of $\vec{S}$ \cite{AlodjantsArakelian1999}. Further, some find it strange that classically unpolarized light can be rotated into an orthogonal state by a transformation that preserves $p$ \cite{Klyshko1992,Bjorketal2010}. Others worry that the two-mode vacuum has $p=1$; or, they define the degree of polarization for the two-mode vacuum as $p\left(\ket{0}_{\hat{a}}\otimes\ket{0}_{\hat{b}}\right)=0$ and dislike the resulting discontinuity in $p\left(\ket{\psi}_{\hat{a}}\otimes\ket{0}_{\hat{b}}\right)$ as $\ket{\psi}_{\hat{a}}\to \ket{0}_{\hat{a}}$ \cite{Klimovetal2005}. A final grief is that some authors want mixtures of perfectly polarized states to have $p\neq 0$ \cite{Klimovetal2005,Bjorketal2010}. These have prompted a plethora of new measures for light's degree of polarization. 

We here include a brief list of solutions to these problems, while maintaining that Eq. (\ref{eq:p degree of polarization}) is a logically consistent measure of degree of polarization that serves as a reference against which all new results are to be measured. We direct the reader to Refs. \cite{Bjorketal2010,Luis2016} for thorough reviews.

Some new degrees of polarization retain use of the Stokes parameters and simply modify Eq. (\ref{eq:p degree of polarization}). They include 
the modification $S_0\to \sqrt{\left\langle\hat{\vec{S}}^2\right\rangle}$,
and agree with Eq. (\ref{eq:p degree of polarization}) in the $S_0\gg 1$ limit \cite{AlodjantsArakelian1999,Klimovetal2010}.
A separate class of definitions for degree of polarization avoids use of the Stokes parameters, instead looking at the aforementioned unpolarized states found without the use of the Stokes parameters \cite{PrakashChandra1971,Agarwal1971}. Luis has pushed to define degree of polarization as the distance between the quasiprobability distributions of a state and of unpolarized light \cite{Luis2002,Luis2007PRAtypeII,Luis2016}. 
In the same vein, other authors define degree of polarization as a function of some distance between a given state and the density operator of unpolarized light. The distance measures include the Hilbert-Schmidt metric, fidelity-based metrics, and metrics based on the trace norm \cite{SanchezSotoetal2006}. Each definition provides a different set of insights into the quantum state under investigation.

There are a number of subsequent definitions of degree of polarization, including entropy-based measures and distinguishability-based measures, with no clear conclusion as to the most viable definition \cite{Luis2007OptComm}. The different definitions give rise to different orderings for which states are more and less polarized than each other \cite{Bjorketal2010}. Nonetheless, Eq. (\ref{eq:p degree of polarization}) is well-defined for all quantum states and provides a canonical ordering for degree of polarization. We retain this global view by seeking to better understand the definition in Eq. (\ref{eq:p degree of polarization}) rather than proposing any particular replacement.

\section{Fixed particle number}
\label{sec:fixed particle number}
\subsection{Perfectly polarized N-photon states: the SU(2) coherent states}
We first look for the $N$-photon states that are perfectly polarized. The most general two-mode $N$-photon pure state can be written as
\eq{
	\ket{\psi^{(N)}}=\sum_{m=0}^{N}c_m\ket{m,N-m},
	\label{eq:general ket}
}
where \eq{\ket{m,n}\equiv\ket{m}_{\hat{a}}\otimes\ket{n}_{\hat{b}}=\frac{\had\vphantom{a}^m\hbd\vphantom{a}^n}{\sqrt{m!n!}} \ket{\text{vac}}} is the state with $m$ photons in mode $\ha$ and $n$ photons in mode $\hb$. The state is ensured to be normalized by $\sum_{m=0}^{N}\left|c_m\right|^2=1$. 

For this state, the first Stokes parameter always satisfies $S_0=N$. We see that the degree of polarization in general obeys $p<1$ despite $\ket{\psi^{(N)}}$ being a pure state. For example, a Fock state with coefficients $c_m=\delta_{mk}$ for some $0\leq k\leq N$ 
has 
perfect polarization $p=1$ only when $k=0$ or $k=N$. The only 2-mode $N$-photon states that are completely polarized and have definite particle number in each of the two modes are states in which one of the two modes is empty.

We next look at the state in which all $N$ photons are in the same mode, associated with the annihilation operator
$
	\hat{a}_{\theta,\phi}=\cos\frac{\theta}{2}\ha+\eu^{-\iu\phi}\sin\frac{\theta}{2}\hb;
$
i.e., we consider the state given by
\eq{
	\ket{\theta\phi^{(N)}}=\frac{\hat{a}_{\theta,\phi}^\dagger\vphantom{a}^N}{\sqrt{N!}}\ket{\text{vac}}.
	\label{eq:theta phi ket}
} A binomial expansion yields
\eq{
	\ket{\theta\phi^{(N)}}
	&=\sum_{m=0}^Nc_m^{\left(\theta,\phi\right)}\ket{m,N-m},
}
where the coefficients are given by
\eq{c_m^{\left(\theta,\phi\right)}=\sqrt{\binom{N}{m}}\cos^m\frac{\theta}{2}\sin^{N-m}\frac{\theta}{2}\eu^{\iu\phi\left(N-m\right)}.
	\label{eq:cm theta phi}
} 
The degree of polarization satisfies $p=1$ for all such states, which are the SU(2) coherent states (or spin-coherent states) \cite{Arecchietal1972}. The SU(2) coherent states are isomorphic to $N$ qubits, to $N$ unit vectors in $\mathds{R}^3$ \cite{Aulbach2011}, and to a spin-$N/2$ particle, if we identify Eq. (\ref{eq:Quantum Stokes operators}) with the Schwinger mapping to angular momentum \cite{Bjorketal2015}. Thus, the SU(2) coherent states comprise all two-mode $N$-photon states that can be written as all $N$ photons being in the same mode, where the latter can be made from a linear combination of the original two modes. All SU(2) coherent states are perfectly polarized.

For the state $\ket{\theta\phi^{(N)}}$, the Stokes vector obeys $\vec{S}/S_0=\left(\cos\theta,\sin\theta\cos\phi,\sin\theta\sin\phi\right)$, which has an obvious interpretation as a unit vector $\hat{n}_{\theta,\phi}$ on the Poincar\'e sphere. 
Using the SU(2) isomorphism, the Poincar\'e sphere is the same as the Bloch sphere for a spin-$N/2$ particle \cite{Migdaletal2014}. Since the operator
\eq{\hat{R}\left(\theta,\phi\right)=\exp\left[\iu\frac{\theta}{2}\left(\hat{S}_2\sin\phi-\hat{S}_3\cos\phi\right)\right]}
rotates a state pointing in the $\hat{z}$ direction to a state pointing in the $\hat{n}_{\theta,\phi}$ direction in the Bloch sphere picture \cite{Arecchietal1972, Migdaletal2014}, it rotates the state $\ket{N,0}$ to the state with all spins pointing in the $\hat{n}_{\theta,\phi}$ direction; i.e., $\hat{R}$ simply rotates the Poincar\'e sphere. This is also seen by identifying the rotated annihilation operator
\eq{\hat{a}_{\theta,\phi}^\dagger=\hat{R}\left(\theta,\phi\right)\hat{a}^\dagger\hat{R}^\dagger\left(\theta,\phi\right),}
so that we can rewrite Eq. (\ref{eq:theta phi ket}) as
\eq{
	\ket{\theta\phi^{(N)}}&=\frac{\hat{R}\left(\theta,\phi\right)\hat{a}^\dagger\vphantom{a}^N\hat{R}^\dagger\left(\theta,\phi\right)}{\sqrt{N!}}\ket{\text{vac}}=\hat{R}\left(\theta,\phi\right)\ket{N,0}
}
(see Appendix \ref{sec:rotation matrix appendix} for an explicit calculation).
There are no other $N$-photon states with perfect polarization, as is quickly seen using the Majorana representation.

\subsection{Majorana representation}
There is a geometrical interpretation of two-mode pure states that lends insight onto interpreting polarization for the states.
The state given by Eq. (\ref{eq:general ket}) can be uniquely rewritten up to a global phase as 
\eq{
	\ket{\psi^{(N)}}=\frac{1}{\sqrt{\mathcal{N}}}\prod_{k=1}^{N}\had_{\theta_k,\phi_k}\ket{\text{vac}},
}
where $\mathcal{N}$ is a normalization factor \cite{Bjorketal2015, Markham2011, Migdaletal2014}. The coefficients $\left\{c_m\right\}$ from Eq. (\ref{eq:general ket}) are found from \eq{c_m=\sqrt{\frac{m!\left(N-m\right)!}{\mathcal{N}}} \prod_{k=1}^{m}\cos\frac{\theta_k}{2}  \prod_{k=m+1}^{N}\eu^{\iu\phi_k}\sin\frac{\theta_k}{2}.  }
The Majorana representation (also known as the Majorana stellar representation) defines a one-to-one mapping from the state $\ket{\psi^{(N)}}$ to the $N$ indistinguishable \cite{Aulbach2011} points $\left\{\left(\theta_m,\phi_m\right)\right\}$ on the Poincar\'e sphere \cite{Bjorketal2015, Markham2011, Migdaletal2014}. It provides a geometrical interpretation for the properties of any $N$-particle two-mode state.

Since $\hat{S}_0$ counts the total number of photons and commutes with the other operators of the algebra, particle number $N$ is conserved by SU(2) operations.
Then any SU(2) operation simply rotates all of the points $\left\{\left(\theta_m,\phi_m\right)\right\}$ about the Poincar\'e sphere \cite{Bjorketal2015, Migdaletal2014}. All of the quantum numbers other than $m$  of the system described by Eq. (\ref{eq:general ket}) are conserved under SU(2) operations \cite{Arecchietal1972}, which encompass all of passive linear optics \cite{Migdaletal2014}. 
Any two states with the same relative orientation, independent of absolute orientation, have the same polarization invariants; changing relative orientation changes polarization invariants \cite{Bjorketal2015}.
This means that the only $N$-particle states with $p=1$, on the surface of the Poincar\'e sphere, are obtained from $\hat{R}\left(\theta,\phi\right)\ket{N,0}$ using an appropriate choice of angles, as claimed above.

To see the power of the Majorana representation, we find the maximum fidelity between an arbitrary $N$-photon state and an SU(2) coherent state. This is calculated by 
\eq{
	\mathcal{F}&=\max_{\theta,\phi}\left|\left\langle{\theta\phi^{(N)}}\big|{\psi^{(N)}}\right\rangle\right|\\
	&=\frac{1}{\sqrt{\mathcal{N}N!}}\max_{\theta,\phi}\left|\bra{\text{vac}}\ha_{\theta,\phi}^N\prod_{k=1}^N\had_{\theta_k,\phi_k}  \ket{\text{vac}}\right|.
}
 {$\mathcal{F}$ is bounded from below by $1/\sqrt{N+1}$ \cite{Martinetal2010}.}
Repeatedly applying the bosonic commutation relations yields the equation
\eq{
	\mathcal{F}&=\sqrt{\frac{N!}{\mathcal{N}}}\max_{\theta,\phi}\left|\prod_{k=1}^N\left(\cos\frac{\theta}{2}\cos\frac{\theta_k}{2}+\sin\frac{\theta}{2}\sin\frac{\theta_k}{2}\eu^{\iu\left(\phi_k-\phi\right)}\right) \right|\\
	&=\sqrt{\frac{N!}{\mathcal{N}}}\max_{\theta,\phi}\prod_{k=1}^N\cos\frac{\Theta_k}{2},
	\label{eq:Max overlap majorana with SU(2)}
}
where $\Theta_k=\cos^{-1}\left(\hat{n}_{\theta,\phi}\cdot\hat{n}_{\theta_k,\phi_k}\right)$ 
is the angle between the unit vectors pointing in the $\left(\theta,\phi\right)$ direction and the $\left(\theta_k,\phi_k\right)$ direction. It is only the relative orientation of the points in the Majorana representation that matter; the average orientation of the points as defined by Eq. (\ref{eq:Max overlap majorana with SU(2)}) yields the closest thing to a perfectly polarized, SU(2) coherent state.

Even after finding the SU(2) coherent state $\ket{\theta\phi^{(N)}}_\text{max}$ that maximally overlaps with $\ket{\psi^{(N)}}$, one is never guaranteed that there exists a constant $\alpha$ such that the state given by
\eq{\ket{\phi^{(N)}}= \frac{\ket{\psi^{(N)}}-\alpha\ket{\theta\phi^{(N)}}_\text{max}}{\sqrt{1-\left|\alpha\right|^2}}}
 is completely unpolarized.
Indeed, unlike in the classical case, an arbitrary $N$-photon pure state cannot in general be decomposed into an SU(2) coherent state pointing in the same direction as $\vec{S}$ and an unpolarized pure state (see Appendix \ref{sec:pure state decomposition}).

We have thus shown that the only perfectly polarized states with fixed particle number are the pure states given by Eq. (\ref{eq:theta phi ket}). The classical decomposition of a state into perfectly polarized and unpolarized components does not work for pure states; such a decomposition will be considered in Section \ref{sec:decomposition into polarized and unpolarized}, following our discussion of mixed states.

\section{Non-fixed particle number}
\label{sec:non fixed particle number}
Next we investigate the perfectly polarized states with indeterminate particle number.
For quantum systems without fixed $N$, the Poincar\'e sphere becomes a series of nested spheres with radii proportional to $N$ \cite{Bjorketal2015}. Each subspace with a given $N$ is decoupled from the other subspaces, and so we can consider an extended Majorana representation with $N$ points on the sphere corresponding to the subspace with $N$ particles. We can break the Stokes vector into Stokes vectors $\vec{S}^{(N)}$ associated with each subspace. SU(2) operations still rotate the Majorana ``constellation"  homogeneously, and leave all polarization invariants unchanged.

\subsection{Pure states}

First we consider the perfectly polarized states with indeterminate particle number that are pure states  .
Given one pure state with $p=1$, all other perfectly polarized pure states are obtained by a solid angle rotation of the state's Majorana constellation. For a particular $N$-particle subspace, it is only possible to achieve perfect polarization with a set of identical Majorana points, so we expect that only superpositions of SU(2) coherent states with different numbers of particles can be perfectly polarized.

Taking the general state given by
\eq{
	\ket{\Psi}=\sum_{N=1}^{M}\eu^{\iu\varphi_N}\sqrt{q_N}\ket{\psi^{(N)}}\quad\sum_{N=1}^{M}q_N=1,
	\label{eq:general ket arbitrary N}
}
we define the Stokes vectors for each $N$-particle subspace by $\vec{S}^{(N)}=\bra{\psi^{(N)}}\hat{\vec{S}}\ket{\psi^{(N)}}$. We find that the degree of polarization, given by the formula
\eq{
	p=\frac{\sqrt{\sum_{N,N^\prime=1}^M \left(q_N\vec{S}^{(N)}\right)\cdot\left(q_{N^\prime}\vec{S}^{(N^\prime)}\right)}}{\sum_{N=1}^M q_NN},
}
relies only on the relative orientations of the Stokes vectors associated with each subspace, and the weight $q_N$ of that subspace in the superposition Eq. (\ref{eq:general ket arbitrary N}). 
This shows that the most polarization is present when the Stokes vectors for all of the subspaces are the most aligned, with the strongest weightings coming from subspaces with the largest populations $q_N\left|\vec{S^{(N)}}\right|$. 

We see by the inequality $\left|\vec{S}^{(N)}\right|\leq N$ that $p$ is less than unity except when the Stokes vectors for each $N$-particle subspace obeys $\left|\vec{S}^{(N)}\right|= N$. Further, $p=1$ requires that all of the vectors $\vec{S}^{(N)}$ point in the same direction. 
Thus, every perfectly polarized pure state can be written as the sum of SU(2) coherent states with different numbers of particles all pointing in the same direction:
\eq{
	\ket{\Psi}_\text{perfect}&=\sum_{N=1}^{M}\eu^{\iu\varphi_N}\sqrt{q_N}\ket{\theta\phi^{(N)}},\quad \sum_{N=1}^M{q_N}=1\\
	&=\sum_{N=1}^{M}\eu^{\iu\varphi_N}\sqrt{q_N}\frac{\had_{\theta,\phi}\vphantom{a}^N}{\sqrt{N!}}\ket{\text{vac}}\\
	&\equiv f\left(\had_{\theta,\phi}\right)\ket{\text{vac}}.
	\label{eq:perfect polarized state fluctuating N}
}
This means that a pure state is perfectly polarized if and only if it can be obtained by the action of an analytic function $f\left(\had_{\theta,\phi}\right)$ of any single creation operator on the vacuum.

In the Majorana representation, the perfectly polarized pure states have the entire constellation aligned on a single ray as it intersects with the surfaces of the nested Poincar\'e spheres. Rotations of the constellation yield all other perfectly polarized pure states. The set of pure states in any mode coupled to the vacuum in the orthogonal mode, and SU(2) rotations of the joint states, yield the entire set of perfectly polarized pure states. 
Again, an arbitrary pure state cannot be decomposed a state of the form Eq. (\ref{eq:perfect polarized state fluctuating N}) pointing in the $\vec{S}$ direction and an unpolarized pure state (see Appendix \ref{sec:pure state decomposition}).

\subsection{Extension to mixed states}
\label{subsec:mixed states}
We have just seen that the most general perfectly polarized pure state is an SU(2) rotation of a pure state in one mode and the vacuum in the orthogonal mode. We now investigate perfectly polarized mixed states given this background.

The known class of perfectly polarized mixed states are those for which an SU(2) rotation diagonalizes one mode and brings the other mode to the vacuum, reminiscent of our pure state result. The former can be written as 
\eq{
	\label{eq:mehta sharma state}
	\hat{\rho}&=\sum_N\rho_N\ket{\theta\phi^{(N)}}\bra{\theta\phi^{(N)}}.
}
This is the original definition of a perfectly polarized mixed quantum state \cite{MehtaSharma1974}.

The density operator formed from $\hat{\rho}_\text{perfect}=\ket{\Psi}_\text{perfect}\bra{\Psi}_\text{perfect}$, corresponding to the property $p=1$, can be brought to the ground state in the $\hat{b}$ mode by the rotation $\hat{R}^\dagger\left(\theta,\phi\right)\hat{\rho}_\text{perfect}\hat{R}\left(\theta,\phi\right)$ as required. However, the resulting state will not be diagonal in mode $\hat{a}$; there are clearly more perfectly polarized states than found in Ref. \cite{MehtaSharma1974}! Indeed, the pure states allowed by Eq. (\ref{eq:mehta sharma state}) are none other than the states $\left\{\ket{\theta\phi^{(N)}}\right\}$, which are but a subset of the pure states defined by Eq. (\ref{eq:perfect polarized state fluctuating N}). We next lay out the true set of conditions required for a state to be perfectly polarized.

\subsubsection{Sufficient condition for perfect polarization}
The definition in Eq. (\ref{eq:mehta sharma state}) is currently cited as comprising the set of perfectly polarized mixed states \cite{SinghPrakash2013,Luis2016}; but, we here show that a less stringent condition is sufficient for a state to be perfectly polarized.
Refs. \cite{PrakashSingh2000,Luis2016}
take Eq. (\ref{eq:mehta sharma state}) as equivalent to the condition 
\eq{
	\ha^{-1}\hb\hat{\rho}\propto \left(\hat{\mathds{1}}_{\ha}-\ket{0}\bra{0}_{\ha}\right)\otimes\hat{\mathds{1}}_{\hb} \hat{\rho}
}
on a perfectly polarized state $\hat{\rho}$,
for $\ha^{-1}=\had\left(1+\had\ha\right)^{-1}$ (see Ref. \cite{Mehtaetal1992} for a discussion of inverse creation and annihilation operators). They extend the condition to pure states with perfect polarization as $\ha^{-1}\hb\ket{\psi}\propto\left(\hat{\mathds{1}}_{\ha}-\ket{0}\bra{0}_{\ha}\right)\otimes\hat{\mathds{1}}_{\hb}\ket{\psi}$, with the unique solution of two-mode \textit{Glauber}
coherent states \cite{TanasKielich1990,Luis2016}. Expressing these solutions in terms of SU(2) coherent states, we have $\ket{G}=\eu^{-r^2/2}\sum_{N=0}^{\infty}\frac{r^N\eu^{\iu\delta N}}{\sqrt{N!}}\ket{\theta\phi^{(N)}}$ for real parameters $r$ and $\delta$ \cite{AtkinsDobson1971}. This is clearly a special case of our solution Eq. (\ref{eq:perfect polarized state fluctuating N}), with $\eu^{\iu\varphi_N}\sqrt{q_N}=r^N\eu^{\iu\delta N-r^2/2}$.

Although it is not readily understood in the Majorana picture, we can extend our generalization to find all of the mixed states with perfect polarization by forming convex combinations of perfectly polarized pure states. Writing the functions
$f_i\left(\had_{\theta,\phi}\right)=\hat{R}\left(\theta,\phi\right)f_i\left(\had\right)\hat{R}^\dagger\left(\theta,\phi\right)$ for
\eq{	
	f_i\left(\had\right)=\sum_N\frac{{\lambda_N^{(i)}}}{{\sqrt{N!}}}\had\vphantom{a}^N,
}
we have the density matrix
\eq{
	\hat{\rho}_{p=1}=\sum_i f_i\left(\had_{\theta,\phi}\right)
	\ket{\text{vac}}\bra{\text{vac}}
	f_i^*\left(\ha_{\theta,\phi}\right).
	\label{eq:perfectly polarized mixed states}
}
We verify $p=1$ by noting that the Stokes parameters obey
\eq{
	S_\mu&=\sum_i\bra{\text{vac}}f_i^*\left(\ha\right)\hat{R}^\dagger\left(\theta,\phi\right)\hat{S}_\mu\hat{R}\left(\theta,\phi\right)f_i\left(\had\right)\ket{\text{vac}}\\
		&=\sum_N\left(\sum_i{\left|\lambda_N^{(i)}\right|^2}\right)\bra{\theta\phi^{(N)}}\hat{S}_\mu\ket{\theta\phi^{(N)}},
}
which is the same as for $\ket{\Psi}_\text{perfect}$ if we write $\left(\sum_i{\left|\lambda_N^{(i)}\right|^2}\right)=q_N$ and $S_\mu^{(N)}=\bra{\theta\phi^{(N)}}\hat{S}_\mu\ket{\theta\phi^{(N)}}$. The most general perfectly polarized state is an arbitrary convex combination of an arbitrary superposition of pure states created in a single, arbitrary mode.

Our main result is that the states in Eq. (\ref{eq:perfectly polarized mixed states}) are equivalent to the density matrices
\eq{\hat{\rho}_{p=1}&=
	\hat{R}\left(\theta,\phi\right)\hat{\sigma}\hat{R}^\dagger\left(\theta,\phi\right)\\
	\hat{\sigma}&=\sum_{N,N^\prime}\sigma_{N,N^\prime}\ket{N}\bra{N^\prime}\otimes \ket{\text{vac}}\bra{\text{vac}}.
} 
This is seen from the fact that the sum $\sum_i \lambda_N^{(i)} \lambda_{N^\prime}^{*(i)}=\sigma_{N,N^\prime}$ can always be formed when $\hat{\rho}$ is Hermitian, given either one of the sets $\left\{\lambda_N^{(i)}\right\}$ or $\left\{\sigma_{N,N^\prime}\right\}$. Alternatively, we can write the density matrices as
\eq{
	\hat{\rho}_{p=1}=\sum_{N,N^\prime}\sigma_{N,N^\prime}\ket{\theta\phi^{(N)}}\bra{\theta\phi^{(N^\prime)}}. 
	\label{eq:perfect mixed state written without functions}}
Thus we have shown the non-trivial result that any general mixed state in one mode combined with the vacuum in an orthogonal mode, and the result of any SU(2) operation acting on the combination, has perfect polarization $p=1$.

\subsubsection{Necessary condition for perfect polarization}
All states of the form Eq. (\ref{eq:perfect mixed state written without functions}) have degree of polarization $p=1$. We now show explicitly that all states with $p=1$ must be of the form Eq. (\ref{eq:perfect mixed state written without functions}).

Consider a general mixed state $\hat{\rho}$ with $p=1$. For such a state, the vector $\vec{S}_{\hat{\rho}}/S_0$ lies on the surface of the Poincar\'e sphere. SU(2) operations affect the orientation of $\vec{S}_{\hat{\rho}}$ but not its length, because this vector transforms under rotations as
\eq{
	\vec{S}_{\hat{\rho}}&\to\text{Tr}\left[\hat{\vec{S}}\left(\hat{R}^\dagger\left(\theta,\phi\right)\hat{\rho}\hat{R}\left(\theta,\phi\right)\right)\right]\\
	&=\text{Tr}\left[\left(\hat{R}\left(\theta,\phi\right)\hat{\vec{S}}\hat{R}^\dagger\left(\theta,\phi\right)\right)\hat{\rho}\right]\\
	&=\hat{R}\left(\theta,\phi\right)\left[\vec{S}_{\hat{\rho}}\right],
}
where $\hat{R}\left(\theta,\phi\right)\left[\vec{x}\right]$ is the rotation operator acting on vector $\vec{x}$.
Consequently, the state $\hat{R}^\dagger\left(\theta,\phi\right)\hat{\rho}\hat{R}\left(\theta,\phi\right)$ also has degree of polarization $p=1$ for all angles $\left(\theta,\phi\right)$, and there can always be found a rotation $\hat{R}\left(\theta,\phi\right)$ such that $\vec{S}_{\hat{R}^\dagger\left(\theta,\phi\right)\hat{\rho}\hat{R}\left(\theta,\phi\right)} =\hat{R}\left(\theta,\phi\right)\left[\vec{S}_{\hat{\rho}}\right]= \left(S_0,0,0\right)$. All mixed states with $p=1$ are obtained from states with $S_0=S_1$ by an SU(2) rotation.

By definition, the first two Stokes parameters satisfy $S_0-S_1\propto\left\langle \hbd\hb \right\rangle$. States with $S_0=S_1$ must then have the vacuum in mode $\hb$.
This leaves complete freedom for mode $\ha$, so all states with $p=1$ are rotated from those with the vacuum in one mode and a general mixed state in the other mode, as in Eq. (\ref{eq:perfect mixed state written without functions}).

The hierarchy is shown in Table \ref{tab:hierarchy table}. Our mixed state result Eq. (\ref{eq:perfect mixed state written without functions}) has the previously accepted Eq. (\ref{eq:mehta sharma state}) as a special case, and the latter allows for the previously accepted pure states Eq. (\ref{eq:theta phi ket}). Our mixed state also allows for the previously unaccepted pure states Eq. (\ref{eq:perfect polarized state fluctuating N}), which again have Eq. (\ref{eq:theta phi ket}) as a special case. This dramatically increases the class of perfectly polarized states.

\begin{table*}[]
	\centering
	\caption{Perfectly polarized states of light. The pure states in each row are special cases of the mixed states in the corresponding rows, and the previously known results in each column are special cases of the new results in the corresponding columns. There are no other states of light with degree of polarization $p=1$, with $p$ defined by Eq. (\ref{eq:p degree of polarization}).}
	\label{tab:hierarchy table}
	\begin{tabular}{|c|c|c|}
		\hline
		\multicolumn{1}{|l|}{} & Pure states & Mixed states \\ \hline
		Previously known & $\ket{\theta\phi^{(N)}}$ & $\sum_{N=0}^\infty\rho_N\ket{\theta\phi^{(N)}}\bra{\theta\phi^{(N)}}$ \\ \hline
		New results & $\sum_{N=0}^{\infty}\sqrt{q_N}\ket{\theta\phi^{(N)}}$ & $\sum_{N,N^\prime=0}^\infty \sigma_{N,N^\prime}\ket{\theta\phi^{(N)}}\bra{\theta\phi^{(N^\prime)}}$ \\ \hline
	\end{tabular}
\end{table*}

\section{Decomposition of arbitrary partially polarized states into perfectly polarized and unpolarized states}
\label{sec:decomposition into polarized and unpolarized}
Now that we have established the full set of states with perfect polarization, we return to the original definition of degree of polarization as quantifying the polarized and unpolarized fractions of a state. If we take our perfectly polarized states from Eq. (\ref{eq:perfectly polarized mixed states}) and combine them with the unpolarized states as defined by Refs. \cite{PrakashChandra1971,Agarwal1971} without reference to the Stokes vectors, we cannot form a basis for all of the two-mode mixed states of light. If, however, we use the original definition of unpolarized light as having $S_1=S_2=S_3=0$, we find that such a decomposition is always possible.

For statistically stationary light, the decomposition $S_\mu=\left(1-p\right)S_{\mu}^{(A)}+pS_\mu^{(B)}$ is unique if we require that the vectorial components of the Stokes parameters obey $\left|\vec{S}^{(A)}\right|=0$ and $\left|\vec{S}^{(B)}\right|=S_0$ \cite{BornWolf1999}. A similar decomposition can be shown to hold for photodector counting rates from beams that are not statistically stationary \cite{LahiriWolf2010}. We can do the same thing for mixed states by writing the density matrix as
\eq{
	\hat{\rho}=\left(1-p\right)\hat{\rho}^{(A)}+p\hat{\rho}^{(B)},
	\label{eq:decomposition of partially polarized mixed state}
}
where $p$ is found using Eq. (\ref{eq:p degree of polarization}). We take $\hat{\rho}^{(B)}$ to be any perfectly polarized state of the form Eq. (\ref{eq:perfect mixed state written without functions}), where the relevant angles are found from the direction of the Stokes vector $\vec{S}=\text{Tr}\left[\hat{\vec{S}}\hat{\rho}\right]$; i.e., we find the unit vector $\hat{n}_{\theta,\phi}=\vec{S}/\left|\vec{S}\right|$. Since the Stokes vector of the perfectly polarized portion is given by
\eq{
	\vec{S}^{(B)}=\text{Tr}\left[\hat{\vec{S}}\hat{\rho}^{(B)}\right]=S_0\hat{n}_{\theta,\phi}=\vec{S}/p,
} 
linearity guarantees that $\vec{S}^{(A)}=\vec{0}$ for all partially polarized states.

The decomposition Eq. (\ref{eq:decomposition of partially polarized mixed state}) is seldom unique. For $p=0$ and $p=1$ there is only one such decomposition, but the variety of perfectly polarized states with the same Stokes vector leads to as many viable decompositions as there are Hermitian matrices for Eq. (\ref{eq:perfect mixed state written without functions}). Perfectly polarized states with fixed particle number are uniquely determined by their Stokes vectors, so specifying the number of particles in $\hat{\rho}^{(B)}$ for a given $\hat{\rho}$ uniquely determines the decomposition Eq. (\ref{eq:decomposition of partially polarized mixed state}).

The full extent of Eq. (\ref{eq:p degree of polarization}) is now understood. The perfectly polarized states of light are those with $p=1$ and are specified by Eq. (\ref{eq:perfect mixed state written without functions}). The completely unpolarized states of light have $p=0$ and are found via Eq. (\ref{eq:decomposition of partially polarized mixed state}) by \eq{\hat{\rho}_\text{unpolarized}=\frac{1}{1-p}\left(\hat{\rho}-p\hat{\rho}_\text{polarized}\right).} Classical light can be uniquely decomposed into a sum of a perfectly polarized and a completely unpolarized beam of light; whereas, the same decomposition for quantum mechanical light is only unique if the perfectly polarized component has fixed particle number. This characterization holds for all two-mode mixed states, regardless of particle number.

\section{Conclusions}
\label{sec:conclusions}
For fixed $N$, the pure states of light with perfect polarization are the SU(2) coherent states, and can be represented by all $N$ Majorana points coinciding on the Poincar\'e sphere. When $N$ is indeterminate, the perfectly polarized states of light are linear combinations of SU(2) coherent states in different subspaces [Eq. (\ref{eq:perfect polarized state fluctuating N})]. These are characterized by $N$ Majorana points at the same place on the Majorana sphere of radius $\propto N$, and the entire Majorana constellation being located on a single ray from the origin. Previous results ignore this class of perfectly polarized pure states.

Earlier results claimed that perfectly polarized mixed states are those obtained by rotating a diagonal density matrix in Fock basis of one mode combined with the ground state of the orthogonal mode; we here show that any density matrix in the first mode will suffice to ensure perfect polarization [Eq. (\ref{eq:perfect mixed state written without functions})]. 

The Stokes parameters can be used to characterize the decomposition of any state into a sum of a perfectly polarized and a completely unpolarized state [Eq. (\ref{eq:decomposition of partially polarized mixed state})]. This sum is not unique for quantum states of light unless the number of particles in the perfectly polarized component is specified.
The class of states that remain unchanged by an ideal polarizer is now seen to include a new set of pure states with varying particle number and a large number of new mixed states.
Perfectly polarized beams are perfectly transmitted through ideal polarizers. A general mixed state incident on an ideal polarizer will transmit only the polarized fraction, but there is no guarantee that this will be a pure state. This is important to realize when working with polarizers in optical circuits.

Definitions for degree of polarization may benefit from this new characterization of the perfectly polarized states. The canonical method of defining degree of polarization is now fully understood and can be used as a benchmark for any future definitions.
The perfectly polarized states discussed here should be used in the widespread applications where extremely precise experimental control is desired. Future work will investigate the use of these states in exceeding classical measurement limits.

\begin{acknowledgements}
	A.G. would like to thank Hudson Pimenta and Jesse Cresswell for fruitful discussions. This work was funded by NSERC.
\end{acknowledgements}

%%%%%%%%%%%%%%%%%%%Begin copied .bbl file
%%%%%%%%%%%%%%%%%%%End copied .bbl file

\onecolumngrid
\begin{appendix}	
\section{Rotation matrix algebra}
\label{sec:rotation matrix appendix}
We here show explicitly that the rotation operator acting on an $N$-photon state with all photons in one mode yields an SU(2) coherent state: $\hat{R}\left(\theta,\phi\right)\ket{N,0}=\ket{\theta\phi^{(N)}}$. Our rotation operator is
\eq{\hat{R}&=\exp\left[\iu\frac{\theta}{2}\left(\sin\phi\hat{S}_{2}-\cos\phi\hat{S}_{3}\right)\right]\\
	&=\exp\left(-\xi\hat{S}_+ +\xi^*\hat{S}_-\right),}
where
$\hat{S}_\pm=\left(\hat{S}_{2}\pm\iu\hat{S}_{3}\right)/2$ and $\xi=\frac{\theta}{2}\exp\left(-\iu\phi\right)$. We follow Ref. \cite{Arecchietal1972} in using the faithful $2\times2$ representation of the $\mathfrak{su}$(2) algebra:
\eq{
	\hat{S}_+=
	\begin{pmatrix}
		0& 1\\0 &0
	\end{pmatrix}
	,\,&
	\hat{S}_-=
	\begin{pmatrix}
		0& 0\\1 &0
	\end{pmatrix}
	,\,&
	\hat{S}_z=
	\begin{pmatrix}
		1& 0\\0 &-1
	\end{pmatrix}
}
to find the $2\times2$ representation of the rotation operator
\eq{
	\hat{R}\left(\theta,\phi\right)=\begin{pmatrix}
		\cos\left|\xi\right|&-\sqrt{\frac{\xi}{\xi^*}}\sin\left|\xi\right|\\
		\sqrt{\frac{\xi^*}{\xi}}\sin\left|\xi\right| &\cos\left|\xi\right|
	\end{pmatrix}=\begin{pmatrix}
		\cos\frac{\theta}{2}&-\eu^{-\iu\phi}\sin\frac{\theta}{2}\\
		\eu^{\iu\phi}\sin\frac{\theta}{2} &\cos\frac{\theta}{2}
	\end{pmatrix}.
}
This is compared to the operator
\eq{
	\hat{Q}\left(a,b,c\right)&=\exp\left(a\hat{S}_-\right)\exp\left(b\hat{S}_z\right)\exp\left(c\hat{S}_+\right)\\
	&=\begin{pmatrix}
		1&0\\a&1
	\end{pmatrix}
	\begin{pmatrix}
		\eu^b&0\\0&\eu^{-b}
	\end{pmatrix}
	\begin{pmatrix}
		1&c\\0&1
	\end{pmatrix}\\
	&=\begin{pmatrix}
		\eu^b&c\eu^b\\a\eu^b&\frac{1+ac\eu^{2b}}{\eu^b}
	\end{pmatrix}.
}
Choosing $a=\eu^{\iu\phi}\tan\frac{\theta}{2}$, $b=\ln\left(\cos\frac{\theta}{2}\right)$, and $c=-\eu^{-\iu\phi}\tan\frac{\theta}{2}$, we find that the operators are related by
\eq{
	\hat{R}\left(\theta,\phi\right)=\hat{Q}\left(a,b,c\right)=\exp\left(\eu^{\iu\phi}\tan\frac{\theta}{2}\hat{S}_-\right)\exp\left(\ln\left(\cos\frac{\theta}{2}\right)\hat{S}_z\right)\exp\left(-\eu^{-\iu\phi}\tan\frac{\theta}{2}\hat{S}_+\right).
}
Because the exponential operators are a faithful representation of the rotation group, this relation is true for all of SU(2).

We can then use the fact that $\hat{S}_+\ket{N,0}=0$ to show explicitly the identity
\eq{
	\hat{R}\left(\theta,\phi\right)\ket{N,0}&=\exp\left(\eu^{\iu\phi}\tan\frac{\theta}{2}\hat{S}_-\right)\exp\left(\ln\left(\cos\frac{\theta}{2}\right)\hat{S}_z\right)\exp\left(-\eu^{-\iu\phi}\tan\frac{\theta}{2}\hat{S}_+\right)\ket{N,0}\\
	&=\exp\left(\eu^{\iu\phi}\tan\frac{\theta}{2}\hat{S}_-\right)\exp\left(\ln\left(\cos\frac{\theta}{2}\right)\hat{S}_z\right)\ket{N,0}\\
	&=\exp\left(\eu^{\iu\phi}\tan\frac{\theta}{2}\hat{S}_-\right)\exp\left(\ln\left(\cos\frac{\theta}{2}\right)N\right)\ket{N,0}\\
	&=\cos^N\frac{\theta}{2}\sum_{n=0}^N\frac{\eu^{\iu n\phi}\tan^n\frac{\theta}{2}}{n!}\hat{S}_-^n\ket{N,0}\\
	&=\cos^N\frac{\theta}{2}\sum_{n=0}^N\frac{\eu^{\iu n\phi}\tan^n\frac{\theta}{2}}{n!}\sqrt{\frac{N!}{\left(N-n\right)!}n!}\ket{N-n,n}\\
	&=\sum_{m=0}^N\sqrt{\binom{N}{m}}\cos^m\frac{\theta}{2}\sin^{N-m}\frac{\theta}{2}\eu^{\iu\phi\left(N-m\right)}\ket{m,N-m}\\
	&=\ket{\theta\phi^{(N)}}
}
as promised.
\section{Pure state decomposition into polarized and unpolarized fractions}
\label{sec:pure state decomposition}
We here show that a pure state cannot in general be decomposed into a pure state perfectly polarized in the direction of $\vec{S}$ and a completely unpolarized state. 

Consider a two-mode pure state $\ket{\Psi}$ and find the angular coordinates  $\Omega=\left(\theta,\phi\right)$ of the vector $\vec{S}=\bra{\Psi}\hat{\vec{S}}\ket{\Psi}$ such that we can define a unit vector by $\hat{n}_\Omega=\vec{S}/\left|\vec{S}\right|$. Then  a general state perfectly polarized in the $\hat{n}_\Omega$ direction is given by the equation
\eq{\ket{\Omega}=\sum_N c_N\ket{\theta\phi^{(N)}}.}
Our task is to show that there does not always exist a state given by
\eq{\ket{\Phi}=\frac{\eu^{\iu\gamma}}{\sqrt{1-\left|\alpha\right|^2}}\left(\ket{\Psi}-\alpha\ket{\Omega}\right)}
for variables $\gamma\in\mathds{R},\,\alpha\in\mathds{C}$ such that the expectation value $\bra{\Phi}\hat{\vec{S}}\ket{\Phi}$ vanishes.

We start by expanding the expectation value to find the relation
\eq{
	\bra{\Phi}\hat{\vec{S}}\ket{\Phi}&\propto 
	\left(\bra{\Psi}-\alpha^*\bra{\Omega}\right)
	\hat{\vec{S}}
	\left(\ket{\Psi}-\alpha\ket{\Omega}\right)\\
	&=\hat{n}_\Omega\left(\left|\vec{S}\right|+\left|\alpha\right|^2\sum_NN\left|c_N\right|^2\right)-2\Re\left[\alpha^*\bra{\Omega}\hat{\vec{S}}\ket{\Psi}\right].
}
$\alpha$ can be tuned to set this expectation value to $\vec{0}$ in specific cases but not in general. We make the decompositions $\alpha=\left|\alpha\right|\eu^{-\iu\beta}$ and $\bra{\Omega}\hat{\vec{S}}\ket{\Psi}=\vec{S}_1+\iu\vec{S}_2$ for real scalar $\beta$ and three-component vectors $\vec{S}_1$ and $\vec{S}_2$ in $\mathds{R}^3$. Then we obtain the expectation value
\eq{
	\bra{\Phi}\hat{\vec{S}}\ket{\Phi}\propto 
	\hat{n}_\Omega\left(\left|\vec{S}\right|+\left|\alpha\right|^2\sum_NN\left|c_N\right|^2\right)-2\left|\alpha\right|\left(\vec{S}_1\cos\beta+\vec{S}_2\sin\beta\right).
	\label{eq:phi expectation value to set to 0}
}
If $\vec{S}_1$ and/or $\vec{S}_2$ point in the direction of $\hat{n}_\Omega$, $\left|\alpha\right|$ and $\beta$ can be tuned to set $\hat{n}_\Omega\left(\left|\vec{S}\right|+\left|\alpha\right|^2\sum_NN\left|c_N\right|^2\right)=2\left|\alpha\right|\left(\vec{S}_1\cos\beta+\vec{S}_2\sin\beta\right)$. If, on the other hand, neither $\vec{S}_1$ nor $\vec{S}_2$ are collinear with $\hat{n}_\Omega$, then the equality $\hat{n}_\Omega\left(\left|\vec{S}\right|+\left|\alpha\right|^2\sum_NN\left|c_N\right|^2\right)=2\left|\alpha\right|\left(\vec{S}_1\cos\beta+\vec{S}_2\sin\beta\right)$ is only possible when $\hat{n}_\Omega$, $ \vec{S}_1$, and $\vec{S}_2$ are coplanar\footnote{Even the coplanar case does not guarantee this possibility. When $\vec{S}_1$ and $\vec{S}_2$ are collinear with each other but not with $\hat{n}_\Omega$, one must find a $\beta$ such that $\vec{S}_1\cos\beta+\vec{S}_2\sin\beta=\vec{0}$, but $\left|\vec{S}\right|+\left|\alpha\right|^2\sum_NN\left|c_N\right|^2\geq 0$ precludes any nontrivial ways of simultaneously setting $\bra{\Phi}\hat{\vec{S}}\ket{\Phi}$ to $\vec{0}$. }.

Since in general $\hat{n}_\Omega$, $\vec{S}_1$, and $\vec{S}_2$ are not coplanar, there does not always exist a state $\ket{\Phi}$ satisfying the desired criteria. For example, the state $\ket{\Psi}=\left(\ket{0,N}+\ket{N-1,1}\right)/\sqrt{2}$ ($N>1$) has Stokes vector $\vec{S}=\left|\vec{S}\right|\left(1,0,0\right)$. Then the most general perfectly polarized pure state in the direction of $\vec{S}$ is $\ket{\Omega}=\sum_Mc_M\ket{M,0}$, and we have the overlap
\eq{
	\bra{\Omega}\hat{\vec{S}}\ket{\Psi}&=\sqrt{\frac{N}{2}}c_N^*\left(0,1,-\iu\right)\\
	&=\sqrt{\frac{N}{2}}\left[\left(0,\Re\left[c_N\right],-\Im\left[c_N\right]\right)+\iu\left(0,-\Im\left[c_N\right],-\Re\left[c_N\right]\right)\right]\\
	&\equiv\vec{S}_1+\iu\vec{S}_2.
}
The three vectors $\hat{n}_\Omega$, $\vec{S}_1$, and $\vec{S}_2$ are far from coplanar; they are orthogonal! Setting $c_N=0$ does not help, because $\left|\vec{S}\right|+\left|\alpha\right|^2\sum_NN\left|c_N\right|^2\geq 0$ in Eq. (\ref{eq:phi expectation value to set to 0}). Therefore one cannot in general decompose a partially polarized pure state into the sum of a pure state perfectly polarized in the direction of $\vec{S}$ and a completely unpolarized state. 

The astute reader will notice that one can always decompose a partially polarized pure state into the sum of a perfectly polarized pure state and a completely unpolarized state if the perfectly polarized pure state is in a direction other than $\hat{n}_\Omega=\vec{S}/\left|\vec{S}\right|$. For example, one can always take the unit vector in the opposite direction to $\vec{S}$ by choosing $\hat{n}_\Omega=-\vec{S}/\left|\vec{S}\right|$, tune the coefficients $\left\{c_N\right\}$ such that $\bra{\Omega}\hat{\vec{S}}\ket{\Psi}=\vec{0}$, and then set $\left|\alpha\right|^2=\left|\vec{S}\right|/\sum_NN\left|c_N\right|^2$. However, this result is not physically useful, as it is strange to decompose a state partially polarized in one direction into a state perfectly polarized in another direction and a state that is completely unpolarized. We thus conclude that the classical decomposition of a partially polarized state into the sum of a state perfectly polarized in that direction and an unpolarized state is not possible for quantum pure states.

\end{appendix}
\end{document}